\documentclass[11pt,a4paper]{article}

\usepackage[margin=1in]{geometry}
\usepackage{amsmath,amssymb,amsthm}
\usepackage{booktabs}
\usepackage{mathrsfs}
\usepackage[hidelinks]{hyperref}

\newcommand{\FF}{\mathbb{F}}
\newcommand{\HH}{\boldsymbol{H}}
\newcommand{\ellr}{\ell_2}

\newtheorem{theorem}{Theorem}

\title{Covering 1024 syndromes with 50 columns}
\author{Stephen Wu}
\date{24 August 2026}

\begin{document}
\maketitle

\begin{abstract}
We exhibit a binary linear $[50,40]_2$ code of covering radius~$2$, so
$\ellr(10,2)\le 50$, one column below the Kaikkonen--Rosendahl
length~$51$ that has stood since 2003 and that still seeds the $R=2$
family of Davydov--Marcugini--Pambianco (arXiv:2511.02542). The new
matrix admits a $(2,0)$-partition into ten blocks, so
Construction~$\mathrm{QM}_2^2$ propagates it to exhaustively verified
codes of lengths $815$ and $1631$ at $r=18$ and $r=20$, and to the
family $n=51\cdot 2^{r/2-5}-1$ of asymptotic density $2601/2048$.
The matrices, verifiers, and \LaTeX\ source are at
\url{https://github.com/wustep/maths}. The pin is
\texttt{problems/covering/share/2026-08-24/} at commit
\texttt{736a38f}. Every number is re-checked from the files in that pin.
\end{abstract}

\begingroup
\renewcommand{\abstractname}{Computation}
\begin{abstract}
The author orchestrated large-language-model agents through Grok Bot.
GPT-5.6 Sol found the $50$-column matrix by simulated annealing on
column sets in $\FF_2^{10}$, starting from the Kaikkonen--Rosendahl
$51$-column set of 2003. Claude Opus~5 found the $10$-block
$(2,0)$-partition of that matrix, implemented Construction
$\mathrm{QM}_2^2$, and wrote two independent verifiers with opposite
algorithms (Python and Rust) that recheck the covering condition and
the partition from the committed matrix files.
\end{abstract}
\endgroup

\section{Definitions and background}

\subsection{Covering radius}

Let $C\le\FF_2^n$ be a binary linear code of codimension $r$, so
$\lvert C\rvert=2^{n-r}$. The \emph{covering radius} of $C$ is the least
$R$ such that every vector of $\FF_2^n$ lies within Hamming distance
$R$ of some codeword. Equivalently, if $\HH$ is an $r\times n$
parity-check matrix of $C$ with column set $S\subset\FF_2^r$, the
covering radius is the least $R$ such that every syndrome is a sum of
at most $R$ columns of $\HH$. Write $\ellr(r,R)$ for the least length
of a binary linear code with codimension $r$ and covering radius $R$.

A binary linear code of length $n$ and dimension $k=n-r$ is an
$[n,k]_2$ code. An $[n,k]_2$ code of covering radius $R$ is written
$[n,k]_2\,R$. For $R=2$ the covering condition on the column set is
\begin{equation}
\label{eq:cover}
\{0\}\cup S\cup(S+S)=\FF_2^r,
\qquad S+S=\{\,s+s' : s,s'\in S\,\}.
\end{equation}

A \emph{$(2,0)$-partition} of the columns (Definition~3.2
of~\cite{DMP2025}) is a partition into nonempty blocks such that every
syndrome, including $0$, is a sum of at most two columns from
\emph{distinct} blocks. Write $p(\HH)$ for the number of blocks of such
a partition.

\subsection{Green's Problem 40}

The covering density of an $[n,n-r]_2\,2$ code is the exact rational
\[
\mu=\frac{1+n+\binom{n}{2}}{2^r},
\]
the number of radius-$2$ representations divided by the number of
syndromes. When \eqref{eq:cover} holds, $\mu\ge 1$. Green's
Problem~40~\cite{Green} concerns the asymptotic quantity
\[
f(2)=\liminf_{r\to\infty}\mu\bigl(\ellr(r,2),r\bigr).
\]
For an infinite family of $[n(r),n(r)-r]_2\,2$ codes, write
$\bar\mu(2)$ for the $\liminf$ of $\mu$ along the family. The optimal
lengths satisfy $\ellr(r,2)\le n(r)$, so any family bound
$\bar\mu(2)\le c$ implies $f(2)\le c$. Whether $f(2)=1$ is open.

The bound $f(2)\le 729/512\approx 1.42383$ held from 1992 to 2025.
Davydov, Marcugini, and Pambianco~\cite{DMP2025} improved it to
$2704/2048\approx 1.32031$ in November 2025. Their tool is a
construction, $\mathrm{QM}_2^2$, that produces an infinite family from
one short radius-$2$ code. The construction applies when the seed
admits a $(2,0)$-partition into few enough blocks.
Section~\ref{sec:prop} states its exact input and output.

\subsection{The seed at $r=10$}

Table~5.1 of~\cite{DMP2025}, row $r=10$, lists $n=51$. The entry is the
Kaikkonen--Rosendahl code of 2003~\cite{KR2003}, of density
$1327/1024\approx 1.29590$. The infinite family of~\cite{DMP2025}
(Theorem~5.7) has lengths
\[
n(r)=26\cdot 2^{r/2-4}-1=2^{m}(51+1)-1,\qquad m=r/2-5 .
\]
The $51$ in this formula is the length-$51$ code; the whole family is
$\mathrm{QM}_2^2$ iterated from that single table entry.
Theorem~5.7(i) of~\cite{DMP2025} calls the $2$-partitions of the seed
``very important for the next iterative process''. Their computer
search gives $p(\HH_{KR})=11$ for the length-$51$ matrix.

A length-$50$ matrix at $r=10$ therefore improves one table entry, and
if it also admits a small $(2,0)$-partition, it re-seeds the whole
family. The matrix of Theorem~\ref{thm:main} and the partition of
Table~\ref{tab:part} do both.

\section{Results}

\subsection{The code}

\begin{theorem}
\label{thm:main}
Let $\HH$ be the $10\times 50$ matrix over $\FF_2$ whose columns are
the $50$ integers of Table~\ref{tab:cols}, where bit $i$ of an integer,
counted from the least significant bit, is row $i+1$. Then $\HH$ has
rank $10$, its columns are distinct and nonzero, and
condition~\eqref{eq:cover} holds. The code with parity-check matrix
$\HH$ is a $[50,40]_2$ code with covering radius exactly $2$, and
\[
\ellr(10,2)\le 50 .
\]
\end{theorem}

The proof is a finite computation over the committed matrix.
Section~\ref{sec:verif} describes the two independent programs that
perform it and the commands that replay it. Appendix~A prints the
matrix as bits.

\begin{table}[ht]
\centering
\small
\begin{tabular}{@{}*{10}{r}@{}}
  1 & 2 & 4 & 15 & 16 & 32 & 65 & 86 & 128 & 173 \\
183 & 202 & 212 & 247 & 256 & 297 & 320 & 329 & 341 & 366 \\
373 & 381 & 391 & 403 & 438 & 460 & 479 & 491 & 502 & 559 \\
576 & 608 & 653 & 734 & 742 & 754 & 771 & 777 & 789 & 821 \\
846 & 855 & 869 & 881 & 893 & 897 & 927 & 981 & 1003 & 1004 \\
\end{tabular}
\caption{The columns of $\HH$ as LSB-first integers.}
\label{tab:cols}
\end{table}

The $1276=1+50+\binom{50}{2}$ sums of at most two columns hit the
$1024$ syndromes with multiplicity histogram
$\{1{:}859,\ 2{:}129,\ 4{:}24,\ 5{:}9,\ 6{:}3\}$. Every multiplicity
is positive, which proves \eqref{eq:cover}. Exactly $973$ syndromes
require two columns, so the covering radius is exactly $2$. The
density is $\mu=1276/1024=319/256=1.24609375$. The columns are distinct
and nonzero, and $\HH$ has a linearly dependent column triple, so the
minimum distance is $d=3$.

\begin{center}
\renewcommand{\arraystretch}{1.15}
\begin{tabular}{@{}ll@{}}
\toprule
claim & value \\
\midrule
shape & $10\times 50$, $\FF_2$-rank $10$, $50$ distinct nonzero columns \\
coverage & $1024/1024$ syndromes \\
covering radius & exactly $2$ \\
density $\mu$ & $319/256=1.24609375$ \\
minimum distance & $d=3$ \\
KR baseline & $51$ columns, $1024/1024$, density $1327/1024$ \\
\bottomrule
\end{tabular}
\end{center}

\subsection{Minimality}

Every matrix obtained from $\HH$ by deleting one column violates
\eqref{eq:cover}: each single deletion leaves at least $9$ of the
$1024$ syndromes uncovered, and the minimum, $9$, is attained by
exactly three columns, $381$, $479$, and $927$. The column set of
$\HH$ is a minimal $1$-saturating set in $PG(9,2)$, and the
covering-code literature calls such a code locally optimal.
Minimality applies to this particular $50$-set. The existence of a
covering $49$-set remains open.

\subsection{A $(2,0)$-partition into ten blocks}

The columns of $\HH$ admit a $(2,0)$-partition with $p(\HH)=10$, shown
in Table~\ref{tab:part}.

\begin{table}[ht]
\centering
\renewcommand{\arraystretch}{1.15}
\begin{tabular}{@{}rl@{}}
\toprule
block & columns \\
\midrule
0 & $2,128,202,212,771,855,897,981$ \\
1 & $86$ \\
2 & $381,893,1003$ \\
3 & $183,297$ \\
4 & $1,65,247,256,320,438,502,734$ \\
5 & $15,173,329,366,460,559,653,846$ \\
6 & $4,16,391,491,742,754,869,881$ \\
7 & $479,1004$ \\
8 & $403$ \\
9 & $32,341,373,576,608,777,789,821,927$ \\
\bottomrule
\end{tabular}
\caption{A $(2,0)$-partition of the columns of $\HH$ with $p(\HH)=10$.}
\label{tab:part}
\end{table}

Each of the $973$ syndromes that need a pair has a pair whose two
columns lie in distinct blocks. Over those $973$ syndromes the pair
multiplicities are $\{1{:}821,\ 2{:}123,\ 4{:}19,\ 5{:}8,\ 6{:}2\}$. In
particular, $821$ syndromes have a unique pair, and each unique pair
forces its two columns into distinct blocks. For comparison, the
computer search of~\cite{DMP2025} gives $p(\HH_{KR})=11$ for the
length-$51$ code. Minimality of the $10$-block partition remains open.

\subsection{Dependent triples}

$\HH$ has exactly $10$ linearly dependent triples of columns. Exactly
one of them, $(491,734,821)$, meets three distinct blocks of the
partition (blocks $6$, $4$, and $9$). This is the analogue of
Theorem~5.2(ii) of~\cite{DMP2025} and is the extra hypothesis that
Construction $\mathrm{QM}_5^2$ needs at $r=28$. That construction is
beyond the scope of this note.

\subsection{Propagation}
\label{sec:prop}

Construction $\mathrm{QM}_2^2$ (Theorem~4.1 of~\cite{DMP2025}) takes an
$[n_0,n_0-r_0]_2\,2$ code whose parity-check matrix $\HH_0$ has a
$(2,0)$-partition into $p(\HH_0)$ blocks, together with an integer $m$
such that $n_0\ge 2^m\ge p(\HH_0)$. In the notation of~\cite{DMP2025},
its indicator set is
$\mathscr{B}=\FF_{2^m}$ and its tail block is
$\boldsymbol{D}=\boldsymbol{D}_1(2)$. The output is a code with
\begin{equation}
\label{eq:qm}
n=2^m(n_0+1)-1,\qquad r=r_0+2m,\qquad p(\HH_C)\le 2^{m+1}+1 .
\end{equation}
With $n_0=50$ and $p(\HH_0)=10$ the hypothesis permits exactly $m=4$
and $m=5$, since $2^6=64>50$. Both outputs come from equations
(4.2) and (4.4) of~\cite{DMP2025} and were verified by enumeration of
every syndrome:

\begin{center}
\renewcommand{\arraystretch}{1.15}
\begin{tabular}{@{}lrrlll@{}}
\toprule
 & $r$ & $n$ & coverage & density & published \\
\midrule
seed & 10 & 50 & $1024/1024$ & $319/256$ & $51$ (KR 2003) \\
$m=4$ & 18 & 815 & $262144/262144$ & $332521/262144$ & $831$ \\
$m=5$ & 20 & 1631 & $1048576/1048576$ & $1330897/1048576$ & $1663$ \\
\bottomrule
\end{tabular}
\end{center}

The indicator assignment inside $\mathrm{QM}_2^2$ is a free choice. A
second assignment gives a second pair of matrices, committed under
\path{result/data/alt/}, and both pairs pass the same verification.

\subsection{The family}

Iterating \eqref{eq:qm} from $(r,p)=(10,10)$, using the partition
bound $p(\HH_C)\le 2^{m+1}+1$ at every step, reaches the even $r\le 64$
accessible by $\mathrm{QM}_2^2$:
\[
10,18,20,30,32,34,36,38,40,46,48,50,52,54,56,58,60,62,64 .
\]
This iteration leaves $r=12,14,16,22,24,26,28,42,44$.
The paper fills its own gaps at $r=22,24,26$ with $\mathrm{QM}_3^2$ and
at $r=28$ with $\mathrm{QM}_5^2$. On the reachable $r$ the lengths and
densities have the closed form
\[
n=51\cdot 2^{r/2-5}-1,\qquad
\mu=\frac{51^2}{2^{11}}-\frac{51}{2^{r/2+1}}+\frac{1}{2^r}
\ \nearrow\ \frac{2601}{2048}\approx 1.27002 .
\]
Hence $\bar\mu(2)\le 2601/2048$, and therefore
$f(2)\le 2601/2048$.

\begin{center}
\renewcommand{\arraystretch}{1.15}
\begin{tabular}{@{}lll@{}}
\toprule
 & $\bar\mu(2)$ & source \\
\midrule
pre-2025 & $729/512\approx 1.42383$ & standing since 1992 \\
arXiv:2511.02542 & $2704/2048\approx 1.32031$ & seeded by KR $n_0=51$ \\
this seed & $2601/2048\approx 1.27002$ & seeded by $n_0=50$ \\
\bottomrule
\end{tabular}
\end{center}

The family rows past $m=5$ follow from the bound
$p(\HH_C)\le 2^{m+1}+1$ of~\cite{DMP2025}.

\section{The search}
\label{sec:search}

\subsection{The matrix}

The $50$-column matrix $\HH$ was found by simulated annealing on
column sets in $\FF_2^{10}$. A state is a set of exactly $50$ distinct
nonzero columns. An array of $1024$ counters records, for each
syndrome, its number of representations among the
$1+50+\binom{50}{2}=1276$ sums of at most two columns. The energy of a
state is its number of zero counters. Replacing one column changes
only that singleton and the $49$ pair-sums that use it, so a swap
costs about $100$ counter updates instead of a $1276$-term recount.
Most proposals target an uncovered syndrome $g$: the search offers $g$
itself, or $g+s$ for a current column $s$.

The published $r=10$ seed is the Kaikkonen--Rosendahl length-$51$
matrix. Before any partition work it was reconstructed from the
hexadecimal listing in Theorem~4.3 of~\cite{DMP2025}. That listing is
MSB-first; the repository stores columns LSB-first
(Section~\ref{sec:encoding}). The reconstructed matrix has $51$
distinct columns and covers $1024/1024$ syndromes, so it is the
published table entry. The search below improves that seed by one
column.

The anneal starts from the best single deletion of this $51$-set:
deleting column $419$ ($\mathtt{0x1A3}$) leaves $11$ syndromes
uncovered. Energy is the number of uncovered syndromes. Moves that
preserve or lower energy always pass; moves that raise it by $\Delta$ pass
with probability $e^{-\Delta/T}$ (Metropolis). With a fixed xorshift64
state, the first run of the script reached a covering $50$-set at
proposal $3{,}600{,}281$. The search stopped there. Partitioning and
Construction $\mathrm{QM}_2^2$ were separate computations.

An independent check of the resulting matrix, from the committed
column text, gives a $10\times 50$ matrix of $\FF_2$-rank $10$ with
$50$ distinct nonzero columns, coverage $1024/1024$, and at least one
syndrome that requires two columns. That finite check is the
one-column improvement of Table~5.1 of~\cite{DMP2025}.

The same method failed at three nearby parameters. At $(r,n)=(8,25)$
the best anneal left $3$ of $256$ syndromes uncovered. At $(9,38)$ it
left $8$ of $512$. At $(10,49)$ it left $7$ of $1024$.

\subsection{The partition}

The published $r=10$ row is the seed of the whole
$\mathrm{QM}_2^2$ family (Theorem~5.7 of~\cite{DMP2025}). A covering
$50$-set is enough for $\ellr(10,2)\le 50$. The family bound needs a
$(2,0)$-partition with few enough blocks. That was a second
computation.

Of the $1024$ syndromes, $973$ need a pair. Their pair-multiplicities
are
\[
\{1{:}821,\ 2{:}123,\ 4{:}19,\ 5{:}8,\ 6{:}2\}.
\]
Each of the $821$ unique-pair syndromes forces its two columns into
distinct blocks. These constraints form a graph with $821$ edges on
the $50$ columns; the maximum degree is $42$. A valid
$(2,0)$-partition is a proper colouring of this graph that, for each
of the remaining $152$ syndromes, also splits at least one of its $2$
to $6$ pairs. Checking a candidate is one pass over the $1276$ sums of
at most two columns.

A colouring with $10$ blocks exists and verifies; it is the artifact
\path{result/data/partition_p10.json}. That committed file certifies
$p(\HH)=10$. The search found colourings with $16$, $12$, and $11$
blocks on the way down. It found no $9$-block or $8$-block colouring
in that run. Ten blocks is smaller than the computer-searched
$p(\HH_{KR})=11$ of~\cite{DMP2025}, and small enough that
$n_0\ge 2^m\ge p(\HH_0)$ holds for both $m=4$ and $m=5$.

The discovery programs that certified the $50$-set, reconstructed the
Kaikkonen--Rosendahl seed, searched for the partition, and enumerated
the $m=4,5$ outputs sit in
\path{problems/covering/share/2026-08-24/compute/}.
The two independent verifiers of Section~\ref{sec:verif} are in
\path{result/verify/}.

\subsection{The check}

Construction $\mathrm{QM}_2^2$ was implemented from (4.2) and (4.4)
of~\cite{DMP2025}. The $m=4$ and $m=5$ outputs were enumerated over
their full syndrome spaces: length $815$ at $r=18$ covers
$262144/262144$ syndromes, density $\mu=1.26847$, against the
published length $831$; length $1631$ at $r=20$ covers its full space,
against the published length $1663$. The two opposite verifiers of
Section~\ref{sec:verif} re-derive the columns from the matrix text and
use the partition file for block labels only. After they agreed, a
third program outside the repository was run once. The covering
radius of $\HH$ is established by this finite verification.

\section{Verification}
\label{sec:verif}

\subsection{Replay}

Clone the public repository, check out the pinned commit, and run
the pipeline from this pin:
\begin{verbatim}
git clone https://github.com/wustep/maths.git
cd maths
git checkout 736a38f
cd problems/covering/share/2026-08-24/result
./run_all.sh
\end{verbatim}
The pipeline needs \texttt{python3} and \texttt{rustc}. After
the checkout it runs offline with those two tools and deterministic
inputs. It executes $74$ named assertions, exits with status $0$, and
prints byte-identical output across runs. A run takes about two
seconds.
Direct links on the pin:
\url{https://github.com/wustep/maths/tree/736a38f/problems/covering/share/2026-08-24/result}
and
\href{https://github.com/wustep/maths/blob/736a38f/problems/covering/share/2026-08-24/result/run_all.sh}{\texttt{result/run\_all.sh}}.

\subsection{Two verifiers with opposite algorithms}

\path{verify/verify.py} is pair-driven: it enumerates all
$\binom{n}{2}$ pairs, marks the syndromes they reach, then sweeps all
$2^r$ values for unmarked ones. \path{verify/verify.rs} is
syndrome-driven: for each of the $2^r$ syndromes $s$ it scans the
columns $h$ and tests membership of $s\oplus h$. It then runs its own
pair loop and reports only if its two verdicts agree. The two programs
also differ in parser and in rank pivot choice, lowest versus highest
set bit, and both use exact rational arithmetic. \path{run_all.sh}
compares their fact dumps line by line.

Both programs derive their results from the matrix text. After they
agreed, a third program kept outside the repository was run once as a
post-hoc comparison and confirmed every value. The committed replay
is \texttt{./run\_all.sh}.

\subsection{Encoding}
\label{sec:encoding}

Every file in the repository is LSB-first: bit $i$ of a column integer
is row $i+1$. The hex listing of $M_{KR}$ in Theorem~4.3
of~\cite{DMP2025} is MSB-first, with row $1$ as the high bit.
Reconstruction of $\HH_{KR}=[\,I_{10}\mid M_{KR}\,]$ therefore reverses
the ten bits of each column. A verifier that misses the reversal fails
on the Kaikkonen--Rosendahl baseline and only there, which makes the
mistake visible. The builder asserts the relation
$h_5+h_{27}+h_{29}=0$ of Theorem~5.2(ii) of~\cite{DMP2025} as a guard.

A reader who uses the MSB-first convention obtains the bit-reversal of
every column. Bit reversal is an invertible linear map on $\FF_2^{10}$,
so every claim in this note is preserved.

\section{Scope}
\label{sec:notclaimed}

\begin{itemize}
\item \emph{Optimality.} The sphere-covering bound gives
$\ellr(10,2)\ge 45$, since $1+n+\binom{n}{2}\ge 1024$ fails at $n=44$
with $991$. Minimality concerns this $50$-set only. A covering
$49$-set may exist.
\item \emph{Nearby parameters.} The searches at $(10,49)$, $(9,38)$,
and $(8,25)$ left $7$, $8$, and $3$ syndromes uncovered, respectively.
\item \emph{Exhaustive checks.} Only the $m=4$ and $m=5$ outputs are
verified exhaustively. The later rows of the family inherit
$p(\HH_C)\le 2^{m+1}+1$ from (4.4) of~\cite{DMP2025}.
\item \emph{Gaps.} The scope excludes
$r\in\{12,14,16,22,24,26,28,42,44\}$.
\item \emph{Asymptotics.} The result is the upper bound
$f(2)\le 2601/2048$. Whether $f(2)=1$ is open.
\item \emph{Priority.} Table~5.1 of~\cite{DMP2025} is ``best as far as
the authors know''. A proof of $\ellr(10,2)\le 50$ may exist in older
proceedings.
\end{itemize}

\section{Files}

Everything named below is in the public repository
\url{https://github.com/wustep/maths}, under
\texttt{problems/covering/share/2026-08-24/}.

The discovery programs are
\path{problems/covering/share/2026-08-24/compute/}
and the six scripts named below.

\begin{center}
\small
\renewcommand{\arraystretch}{1.15}
\begin{tabular}{@{}ll@{}}
\toprule
path & contents \\
\midrule
\path{result/} & the verification pipeline and matrices \\
\path{result/run_all.sh} & the replay script \\
\path{result/data/H_r10_n50.txt} & the matrix \\
\path{result/data/partition_p10.json} & the $10$-block partition \\
\path{compute/} & 16 August compute plus the discovery programs \\
\path{compute/verify\_H\_r10\_n50\_independent.py} & independent check of $\HH$ \\
\path{compute/reconstruct\_KR\_r10\_n51.py} & Kaikkonen--Rosendahl reconstruction \\
\path{compute/search\_partition\_p10.py} & partition search \\
\path{compute/verify\_partition\_p10.py} & partition check \\
\path{compute/propagate\_qm\_m4.py} & $\mathrm{QM}_2^2$ at $m=4$ \\
\path{compute/propagate\_qm\_m5.py} & $\mathrm{QM}_2^2$ at $m=5$ \\
\bottomrule
\end{tabular}
\end{center}

\appendix
\section{The matrix as bits}

Rows $1$ to $10$ from top to bottom, columns left to right. Bit $i$
(LSB) of each integer column is row $i+1$.

{\small
\begin{verbatim}
10010010011001010110111100110100100011110111111110
01010001001101000001001110111100011110001100001010
00110001011011000011111011101100111000111110101101
00010000010100010101010001110100110001001000101011
00001001001011000010110110101000010100110101101100
00000100011001010001110010011101001100010011100011
00000011000111001111110001111011011100001111100111
00000000111111000000001111111000111100000000011111
00000000000000111111111111111000000011111111111111
00000000000000000000000000000111111111111111111111
\end{verbatim}
}


\begin{thebibliography}{9}
\bibitem{DMP2025}
A.~A. Davydov, S.~Marcugini, F.~Pambianco,
\emph{New upper bounds for binary linear covering codes},
arXiv:2511.02542 [cs.IT], November 2025.

\bibitem{KR2003}
M.~Kaikkonen, C.~Rosendahl,
\emph{New covering codes from an ADS-like construction},
IEEE Trans.\ Inform.\ Theory \textbf{49} (2003), no.~7, 1809--1812.

\bibitem{CHLL}
G.~Cohen, I.~Honkala, S.~Litsyn, A.~Lobstein,
\emph{Covering Codes}, North-Holland Mathematical Library \textbf{54},
Elsevier, Amsterdam, 1997.

\bibitem{Green}
B.~Green, \emph{100 open problems}, Problem~40.
\url{https://people.maths.ox.ac.uk/greenbj/papers/open-problems.pdf}.
\end{thebibliography}
\end{document}